\newif\ifdraft\drafttrue
\newif\ifcolor\colortrue
\newif\iftechreport\techreporttrue

\iftechreport
\documentclass[onecolumn]{article}
\pdfoutput=1
\usepackage{epsfig}
\usepackage{amssymb}
\usepackage{amsmath}
\usepackage{amsfonts}
\usepackage{amsthm}
\usepackage[margin=1.0in]{geometry}
\else
\documentclass[letter]{sig-alternate-10pt}
\fi

\newtheorem{definition}{Definition}

\usepackage[table]{xcolor}
\definecolor{pennblue}{cmyk}{1,.65,0,.3}
\definecolor{pennred}{cmyk}{0,1,.65,.34}

\definecolor{williamspurple}{RGB}{89,23,128}
\definecolor{williamsgold}{RGB}{255,213,0}

\usepackage{arydshln}
\usepackage{booktabs}
\usepackage{cmtt}

\usepackage{paralist}
\usepackage{lastpage}
\usepackage{times}
\usepackage{tikz}
\usepackage{alltt}
\usepackage{graphicx}
\usepackage{balance}
\usepackage{amsmath}
\usepackage{caption}
\usepackage{subcaption}
\usepackage{algorithm,algpseudocode}
\usepackage{enumitem}

\usepackage{verbatim}

\usepackage{tikz}
\pgfdeclarelayer{background}
\pgfdeclarelayer{foreground}
\pgfsetlayers{foreground,background,main}

\iftechreport
\else
\fi
\usepackage[export]{adjustbox}
\graphicspath{{./figures/}}

\begin{document}

\iftechreport
\title{A3: An Automatic Topology-Aware Malfunction Detection and Fixation System in Data Center Networks}
\else
\title{XXX}
\fi

\iftechreport
\author{ Che Zhang \\ SUSTech
\and Shiwei Zhang \\ HKU
\and Bo Jin \\ SUSTech
\and Weichao Li \\ SUSTech
\and Zhen Wang
\and Qing Li \\ SUSTech
\and Yi Wang \\ SUSTech
}

\else
\numberofauthors{1}
\author{Paper \#XXX, \pageref{ConcPage} pages}

\fi

\iftechreport
\date{}
\else
\date{}
\fi

\maketitle

\begin{abstract}
  Link failures and cable miswirings are not uncommon in building data center networks, which prevents the existing automatic address configuration methods from functioning correctly. However, accurately detecting such malfunctions is not an easy task because there could be no observable node degree changes. 
  Fixing or correcting such malfunctions is even harder as almost no work can provide accurate fixation suggestions now.

  To solve the problems, we design and implement A3, an automatic topology-aware malfunction detection and fixation system. A3 innovatively formulates the problem of finding minimal fixation to the problem of computing minimum graph difference (NP-hard), and solves it in $O(k^6)$ and $O(k^3)$ for any less than $k/2$ and $k/4$ undirected link malfunctions for FatTree, respectively. Our evaluation demonstrates that for less than $k/2$ undirected link malfunctions, A3 is 100\% accurate for malfunction detection and provides the minimum fixation result.
   For greater or equal to $k/2$ undirected link malfunctions, A3 still has accuracy of about 100\% and provides the near optimal fixation result.

\end{abstract}

\newcommand{\jnfpar}[1]{\ \\[-.75em] \textit{#1.}~}

\section{Introduction}
\label{sec:introduction}

The process of building a DC is very complex and error-prone.
Existing method \cite{ref:dac,ref:jupiter}, assumes that prior to deploying a real data center,
a blueprint should be designed to guide construction of data center. In a blueprint, interconnections between devices and logical ID for
each device (e.g.,ip encoded with information of topology and locality)  are provided.
Then building a data center often needs to connect thousands of devices with strict wiring rules defined in a blueprint (e.g., FatTree). Malfunctions like device, link failures and miswired cables due to hardware or cabling errors are nearly inevitable. Although in most cases malfunctions cause detectable node degree change, it is still possible to have miswirings with no node degree change.

  Accurately detecting and fixing such malfunctions especially miswirings often requires significant human efforts. It can cause substantial operation delay of the whole data center as recent works either can only auto-configure addresses of devices without malfunctions \cite{ref:dac} or need to know which devices are not involved in malfunctions in order to configure their addresses and let them work first  \cite{ref:etac}.

Therefore, it is very important and urgent to design a system to automatically detect the malfunctions and provide fixation suggestion better to be minimum fixation which means to fix the malfunctions in the least steps. For example, disconnecting (1,6) and (5,4) then connect (1,5) and (4,6) of the last physical graph shown in Figure~\ref{fig:a3prob1} to correct the wrong connections and make it just the same as the blueprint.

DAC \cite{ref:dac} is generic for various topologies and can identify whether there are malfunctions in a data center
by comparising the collected physical graph to a standard blueprint like FatTree by solving a graph isomorphism problem. However, DAC's malfunction detection algorithm can only report a list of possible malfunctions, because it is a heuristic trying to solve an NP-hard maximum common subgraph problem. As DAC's reported malfunctions are not accurate (experiment result is shown in Figure 11 and Figure 12 and the reason of its low accuracy is explained in Section 6.3.2), administrators still need a long time to go through the list and rectify the malfunctions. Moreover, DAC can not provide fixation suggestion  automatically.

We propose A3, an automatic topology-aware data center malfunction detection, fixation and configuration system that addresses the above challenges. We design two exact algorithms and they both can solve the above three problems, malfunction localization, finding mimimum fixation and address auto-configuration for FatTree data center with less than $k/2$ and $k/4$  undirected link malfunctions, respectively (e.g., link failures, cabling errors, miswirings with no node degree change, etc.). For greater or equal to $k/2$ undirected link malfunctions, exact algorithm 1 can still provide feasible fixation in $O(k^6)$. exact algorithm 2 has time complexity $O(k^3)$ and for greater or equal to $k/4$ undirected link malfunctions, it will report the number of malfunctions beyond its bound then algorithm 2 can be triggered or A3 can run both algorithms at the same time. Experiment shows that compared with DAC's fast malfunction detection with at most 12\% accuracy and no fixation suggestion, A3 has 100\% accuracy and provides minimum fixation (both algorithms) with in each algorithm's malfunction bound, and near 100\% accuracy and provides feasible fixation (computed by algorithm 1) for the malfunctions beyond algorithm 1's bound ($k/2$). A3 only uses a slightly longer time for algorithm 1 than DAC's malfunction detection algorithm (with 3\% nodes as anchor points). The poster version is \cite{ref:a3poster} and our source code is \textcolor{blue}{https://github.com/sigcomm2019a3/code}.

\begin{figure}
\centering
\includegraphics[width=0.7\linewidth]{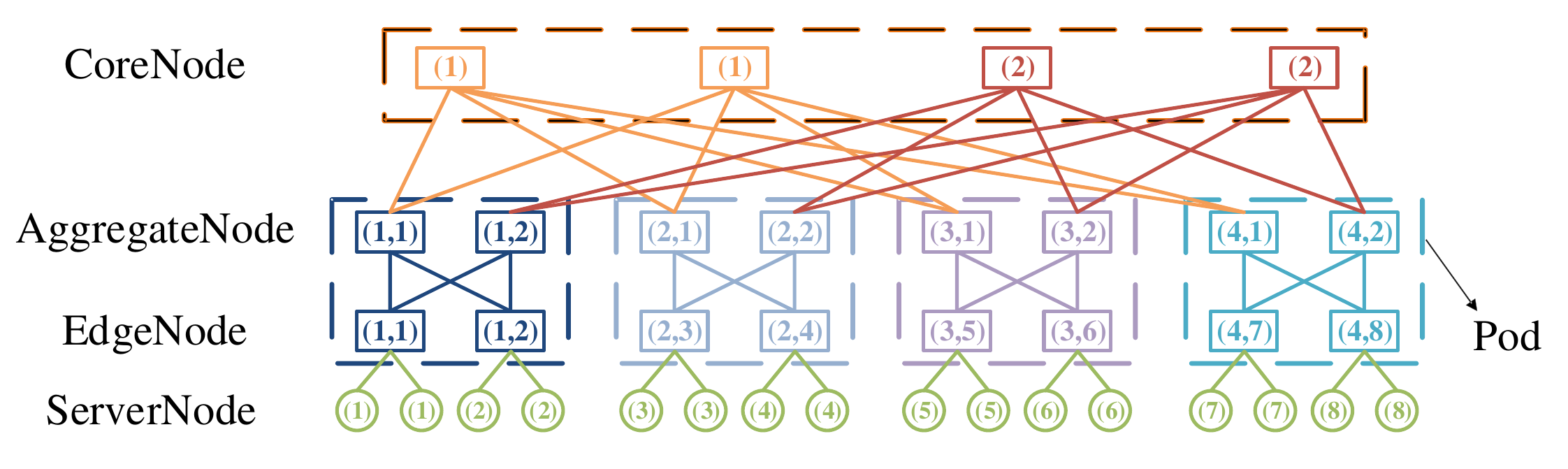}
\vspace{-0.2cm}
\caption{Roles and their connection rules of FatTree(k=4). Each node has a role, e.g., the role of upper left node in the first pod is represented by AggregateNode(group=1,index=1). The color of each edge stands for the connection rules, e.g., when CoreNode's group id equals AggregateNode's index, connecting them. }
\label{fig:colorfattree}
\end{figure}

\begin{figure}
\centering
\vspace{-0.4cm}
\includegraphics[width=0.7\linewidth]{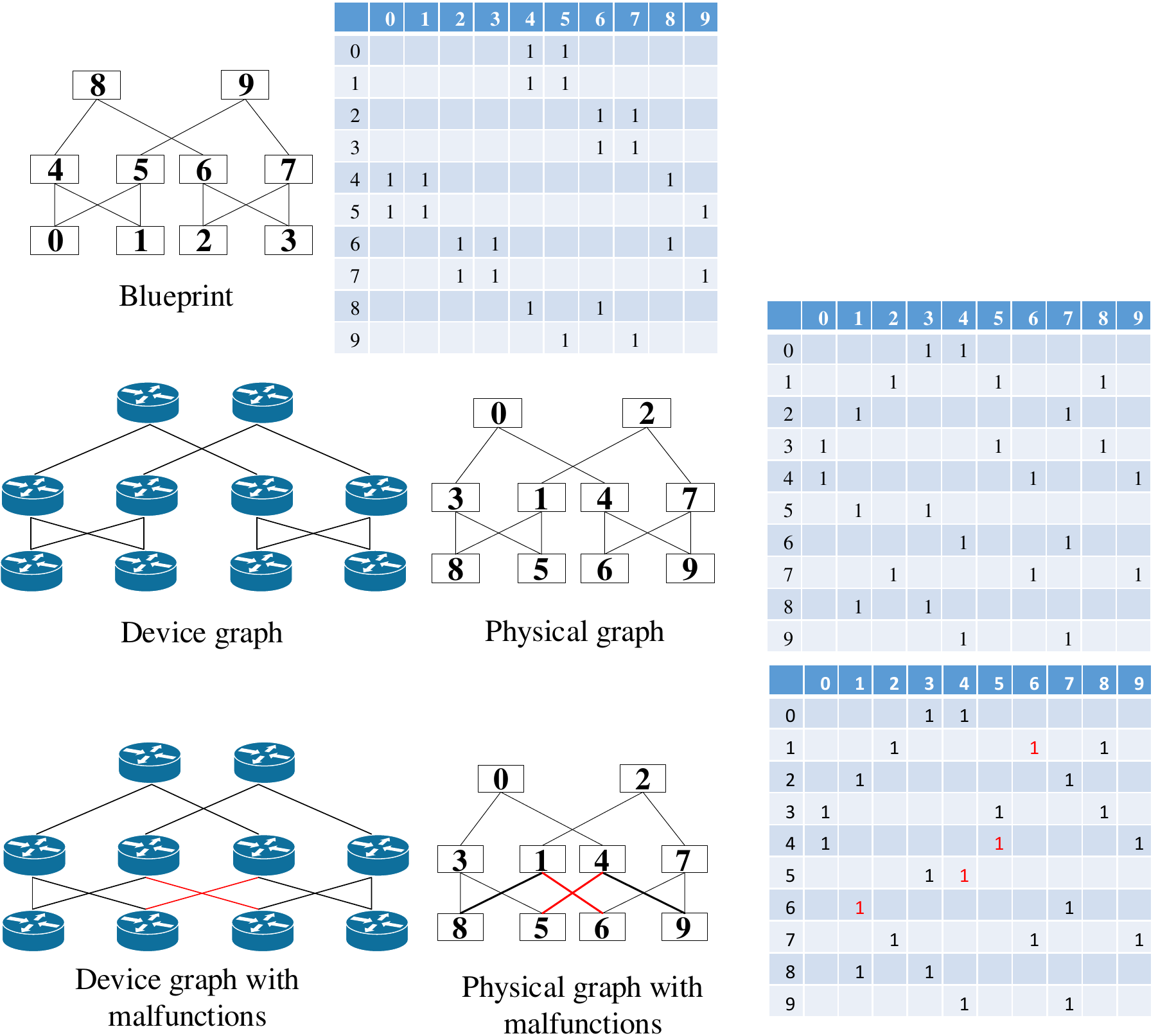}
\vspace{-0.2cm}
\caption[]{Although the upper first two matrices seems not the same, the graphs actually
are isomorphic and the permutation is 0123456789 to 8569314702. However, in reality, cable mis-wirings, etc. are not uncommon in building data center networks. We can not find a permutation to make the blueprint and the third device graph with malfunctions have the same matrix (nonisomorphic).}
\label{fig:a3prob1}
\vspace{-0.1cm}
\end{figure}

The basic idea of A3 is as follows. Instead of designing a blueprint with logical ID, we choose to redesign the blueprint by abstracting the structural property (roles and their connection rules shown in Figure~\ref{fig:colorfattree} and explained in Section 2.2 and 4) which depends on the network structure alone and not, for example, on node or edge labels \cite{ref:ruben}. Then instead of using graph isomorphism algorithm to mapping physical graph to blueprint in order to auto-configure the addresses of devices, or trying to solve maximum common edge subgraph problem (NP-complete) to detect malfunctions, we select to find a labelling of the blueprint (optimal labelling) whose corresponding adjacency matrix has the minimum differences (minimum fixation) compared with the adjacency matrix of the physical graph with malfunctions. Finally, instead of randomly relabelling the blueprint and comparing with the physical graph to find the optimal one, we choose to generate the optimal labelling from the physical graph with malfunctions  directly utilizing the fact that 1.  malfunctions is mostly only a small portion and 2. most data center networks like FatTree are highly symmetric.

As we have abstract the roles and connection rules in the blueprint which means each node has a role, so generating the optimal labelling from the physical graph actually transforms to generating the optimal role assignment from the physical graph (device-to-role mapping). The basic of exact algorithm 1 is greedy algorithm and the basic of exact algorithm 2 is constructing method using roles and connection rules. As you can notice in Figure~\ref{fig:colorfattree}, actually, the role of each node has the topological information that logical ID encodes, so using the device-to-role mapping result, we can get the device-to-logical ID mapping directly (address auto-configuration).

\section{Background}
\label{sec:background}

We begin by discussing the correlation among graph problems of malfunction detection and address configuration. We then motivate the need for a new formulation of malfunction fixation which can also cover them all and prove its NP-hardness. Next, we explain the intuitions that we use the topological property of network to accelerate the solving of our formulation instead of sacrifice the accuracy to solving it generally.

\subsection{Problem Formulation}

Related works~\cite{ref:dac,ref:etac} focus on solving the address configuration problem generally. Their logic is first find the mapping between blueprint and physical graph (\textit{graph isomorphism}
), if failed, trigger the running of malfunction detection and output the list of possible malfunctions (\textit{maximum common subgraph isomorhpism}, NP-hard). Then network managers check the list and localize the real malfunctions and trigger the error tolerent address configuration process~\cite{ref:etac} (\textit{induced subgraph isomorphism}, NP-hard) or try to fix the malfunctions manually before computing the address configuration of the whole network~\cite{ref:dac}.

Such logic of an address configuration system seems reasonable. However, they do not consider duplication computations in each process especially most processes can be formulated as NP-hard problems.
It means, in the worst case that a built data center has malfunctions, three processes are triggered, so the generally used graph mapping operations for solving those problems are duplicated three times.

Therefore, a new formulation is need to cover those processes all in order to reduce such duplication. Fortunately, we formulate the problem of finding minimal fixation to the Minimum Graph Difference Problem (MGDP)
which covers them all.
Given two graphs $G_1=(V_1,E_1)$ and $G_2=(V_2,E_2)$, where $V_1$, $V_2$ are the vertex sets and $E_1$, $E_2$ are the edge sets.
Without loss of generality, we assume $|V_1|=|V_2|$ (otherwise, we can add extra edge-free vertices to $G_1$ or $G_2$ to ensure $|V_1|=|V_2|$).
In addition, let $A_1=[a_1(i,j)\in\{0,1\}]_{V_1\times V_1}$ and $A_2=[a_2(i,j)\in\{0,1\}]_{V_2\times V_2}$ be the adjacency matrices.
The Minimum Graph Difference Problem (MGDP) is the problem of finding a bijection $\pi: V_1\to V_2$, so as to minimize the difference
$d(\pi)=\sum_{(u,v)\in V_1\times V_1} |a_1(u,v) - a_2(\pi(u),\pi(v))|$.

The Maximum Common Edge Subgraph Problem (MCESP) is the problem of finding a graph $H$ with as many edges as possible which is isomorphic
to both a subgraph of $G_1$ and a subgraph of $G_2$, i.e., finding a bijection $f: V_1\to V_2$ such that the number of common edges
$c(\pi)=\sum_{(u,v)\in E_1} 1-|a_1(u,v) - a_2(\pi(u),\pi(v))|$ is maximized. The equivalence between MGDP and MCESP is
revealed by
$d(\pi) + 2\cdot c(\pi) = |E_1| + |E_2|$.
Next, we prove the NP-completeness of MGDP by showing the MCESP is NP-complete. To prove MCESP is NP-complete, it must be formulated as
a decision problem. The input to the decision problem is a pair of graphs $(G_1, G_2)$ and a number $k$. The answer to the problem is positive
if there exists a common subgraph $H$ of both $G_1$ and $G_2$ such that the number of edges in $H$ is at least $k$, and negative otherwise.

The proof of MCESP being NP-complete is
based on reduction of the Subgraph Isomorphism Problem (SIP), which is a well-known NP-complete
problem. The SIP is the problem of answering whether a graph $H$ is isomorphic to a subgraph of $G$. Any SIP instance $(G, H)$ can be reduced to
an MCESP instance with a pair of graphs $(G, H)$ and a number $|H|$. Then the answer to the MCESP instance $(G,H,|H|)$ is equal to the answer of
the SIP instance $(G,H)$.

\subsection{Topological Properties}

We first explain why we use topological properties of network to accelerate the solving of MGDP.

Although there are many graph algorithms have been developed to solve the above problems, \cite{ref:ceci,ref:subisodp} for subgraph or induced subgraph isomorphism,
\cite{ref:mces2012,ref:mcs2017,ref:between} for maximum common edge or induced subgraph isomophism,
etc.,
they are all designed for general graphs which are not scalable for large scale data center networks. For example, \cite{ref:mces2012} mentioned that the graphs they could deal with are limited to less than 40 vertices. \cite{ref:dac,ref:etac} utilize the symmetric property of the data center networks, however, as they are still focused on the generality, the malfunction detection algorithm~\cite{ref:dac} they used sacrifices the accuracy to trade off for much faster results.

We argue that although such sacrifice may be valuable for some graph pattern mining tasks like frequent subgraph mining~\cite{ref:asap} which are often not necessary to output the exact answer, for the problems related to network malfunction detection and malfunction fixation, such sacrifice of accuracy is not worthy and even harmful. The reason is that the network managers still need to check and make clear about the real malfunctions manually even if the system quickly outputs the list of possible malfunctions. It may cost larger human labor if such list is inaccurate and substantial longer operation delays of the whole data center. A typical example is that allowing traffic to use a miscabled link can lead to forwarding loop~\cite{ref:jupiter}.

Moreover, most networks have blueprints to direct the network building process, we can fully exploit the topological properties of blueprints to accelerate the solving of MGDP by sacrificing generality. Such topological properities can also be abstracted to simplify the network building process and reduce the possibility of malfunction occurrence. To guarantee accuracy and efficiency, we can also limit the number of malfunctions the algorithm covers.

We first focus on FatTree, then briefly review key definitions needed to understand A3.

\noindent \textbf{\textit{FatTree introduction:}}
FatTree~\cite{ref:fattree} is a
wildly used topology in DCs
(Figure~\ref{fig:colorfattree}).
A FatTree contains $k^3/4$ servers and $5k^2/4$
$k$-port switches. It is divided into four layers, including server layer, edge layer, aggregation layer and core layer.
And it has $k$ pods, with each pod having $k^2/4$ servers.

The automorphisms of $G$ are the permutations of $V(G)$ that can be applied to both the rows and the columns of $A(G)$ without changing $A(G)$.
As vertices in the same orbit are structurally indistinguishable, orbits contribute to network redundancy and related to the robustness of the underlying system \cite{ref:ruben}. A3's \textbf{roles} actually belong to vertex orbits and nodes which have the same role means they are not only in the same orbit but also connected to the same nodes. FatTree has four vertex orbits, including CoreNodes orbit, AggregateNodes orbit, EdgeNodes orbit and ServerNodes orbit \cite{ref:cheiccc}. Therefore, each role can be represented by the orbit it belongs to and the specific parameters related to its neighbors. Similarly, A3's connection \textbf{rules} among roles belong to edge orbits and the relationship of each rule with the two roles it connected is obtained by the same corresponding parameter of roles.

\noindent \textbf{\textit{Structural network measure}} \cite{ref:ruben}:
A (pairwise) structural network measure
is a function F(i,j) on pairs of vertices which depends on the network structure alone.
Crucially, structural measures are independent of the ordering or labelling of the vertices and hence satisfy, for any automorphism $\delta \in Aut(G)$, $F(\delta(i), \delta(i))=F(i,j)$ for all $i,j \in V$.

\section{Design Overview}
\label{sec:design}

\begin{figure}
\centering
\includegraphics[width=0.7\linewidth]{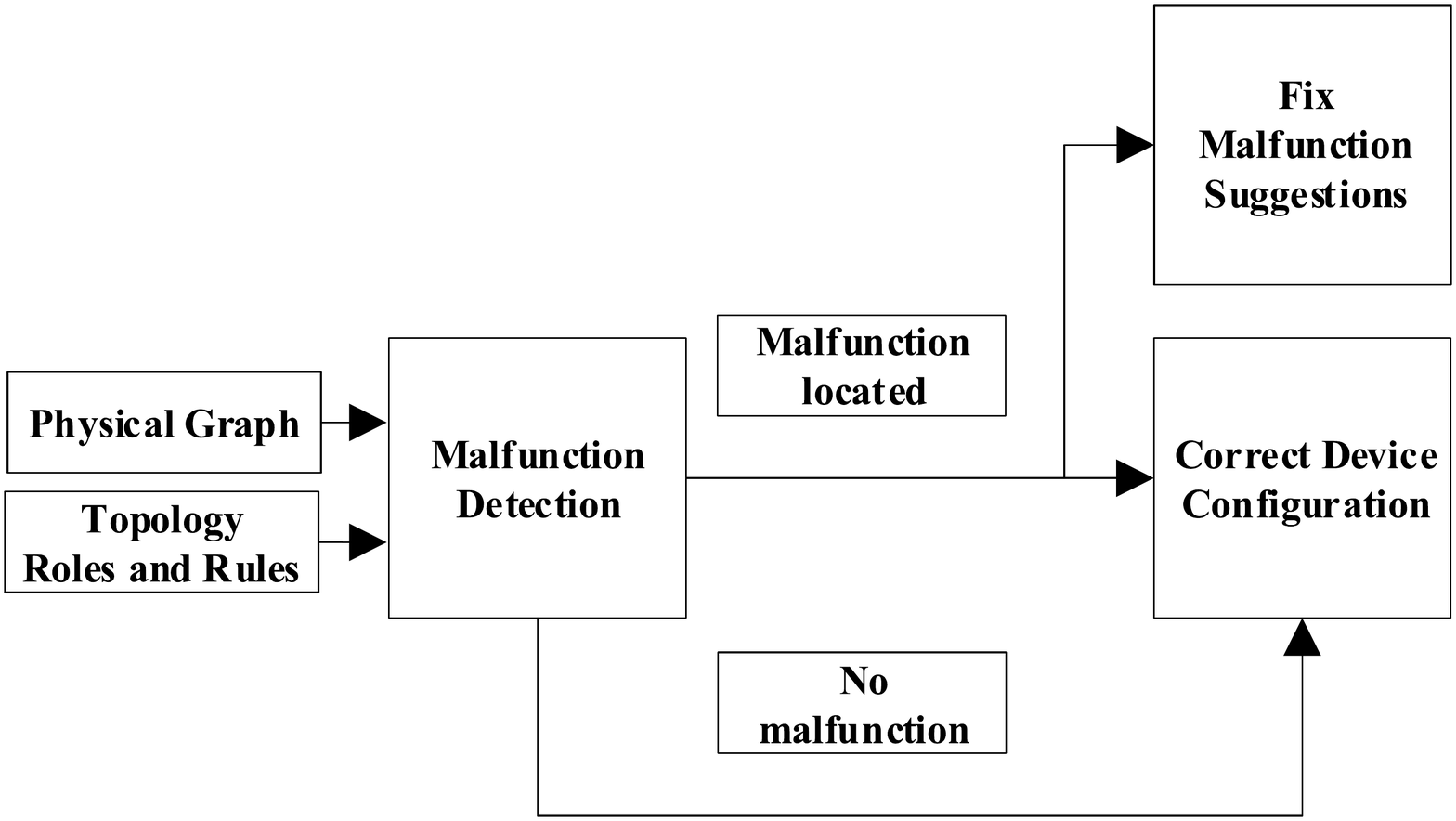}
\caption{The framework of A3.}
\label{fig:system}
\end{figure}
We design A3, a system that facilitates accurate and scalable data center malfunction detection, fixation and configuration by solving MGDP which covers them all.

\noindent \textbf{A3 architecture:} Utilizing the abstracted topological properties, A3 can realize incremental deployment and real time malfunction detection for each connected link to accelerate the building of a data center network. After finishing the building process, physical graph can be collected. As shown in Figure~\ref{fig:system}, A3 only needs the physical graph and topology roles and rules as the input of MGDP.
First, A3 detects if there is any malfunction in the graph. If not, A3 autoconfigures the logical ID of each device as before. If the malfunctions exist,
A3 first locate the malfunctions and then find the minimal fixation suggestion and autoconfigure the correct devices.

Our design logic is that FatTree is very common, therefore, we focus on FatTree first and other FatTree like topologies (e.g., VL2 \cite{ref:vl2} or
Google Jupiter \cite{ref:jupiter}) can also use A3 once their topology roles and rules are extracted. In reality, malfunctions are seldom,
so we design two exact algorithms.

As far as we are concerned, we are the first to put forward that we can use the automorphism in FatTree to
accelerate malfunction detection and then correct the false connections as well as reach the goal of data center address configuration. Up to now, DAC introduced the graph isomorphism to device-to-logical mapping as well as malfunction detection, and used some general characteristics of DCs to solve such a
NP-hard maximum common subgraph problem. We are so inspired to find out some internal features of the extensively used DC structures that is really matters (roles and their connection rules for our re-designed blueprint). Therefore, we introduce the automorphism
and make the NP-hard MGDP problem solvable in P under the limited small number of malfunctions using this characteristic.

\noindent \textbf{Collecting physical graph:} The physical topology is constructed by first putting the type of devices of each role to corresponding location based on role-to-location table. Then
connect the devices based on connection rules. We follow Google Jupiter's design \cite{ref:jupiter}, collecting the physical network topology and
disseminating logical ID using central controller.
Moreover, our A3 can collect the corresponding port-to-port connection information instead of getting it manually or design it in the blueprint.
Therefore, combining with the role information for each device computed by A3, we can also use Google Jupiter's Neighbor Discovery protocol \cite{ref:jupiter} to detect failures for the running data center.

\section{Re-design Blueprint}
\label{redesign}

Building a data center needs two kinds of basic information. 1. Location-to-type of device which means the device type putting in each location.
2. Location-to-location connections (so that builders can know how to connect the devices putting in those locations). We argue that such
location-to-location connection information is the natural requirement for data center cable bundling in order to reduce the fiber cost and
optimize deployment \cite{ref:jupiter}.
DCs have well-defined structure. Prior to deploying a real DC, a blueprint should be designed to guide the construction of
the DC. To make DC construction simple and understandable, we only require the blueprint to provide the following minimal
information.

\noindent \textbf{Role-to-location:} The main purpose of role-to-location table is that it is an easy way to get the location-to-location connections and builders can also check whether the location is right by the label of those locations simply. 

$CoreNode=\{y_1,y_2,...,y_{k^2 \over 4}\}$,

$AggregateNode=\{x_1,x_2,...,x_{k^2 \over 2}\}$,

$EdgeNode=\{w_1,w_2,...,w_{k^2 \over 2}\}$,

$ServerNode=\{v_1,v_2,...,v_{k^3 \over 4}\}$.

$F(v_i,w_j)=1$, if $\lceil i/({k \over 2}) \rceil == j$;

$F(w_i,x_j)=1$, if $\lceil i/({k \over 2}) \rceil == \lceil j/({k \over 2}) \rceil $;

$F(x_i,y_j)=1$, if $i \% ({k \over 2}) == \lceil j/({k \over 2}) \rceil $;

Otherwise, $F(i,j)=0$.

The upper structural network measure can be further expressed and simplified in the following way. The roles  can be generated automatically.

\begin{itemize}
\item $CoreNode(group)$ is the role for core switches. Based on the same connections, core switches can be divided into $k/2$ groups and each group has $k/2$ core switches.
\item $AggregateNode(group,index)$ is the role for aggregation switches. Based on the pods, aggregation switches can be divided into $k$ groups. Each group has $k/2$ aggregation switches and their indexes are the same as the connected core group id.
\item $EdgeNode(group,index)$ is the role for edge switches. Based on the pods, edge switches can be divided into $k$ groups and their group ids are the same as the connected aggregate group id (pod id). Each group has $k/2$ edge switches.
\item $ServerNode(group)$ is the role for servers. Their group id are the same as the connected edge index id. Therefore, servers can be divided into $k^2/2$ groups and each group has $k/2$ servers.
\end{itemize}

It should define the correspondence between roles and their location so that the builders do not need to record the device-to-location information.
Builders build a data center just using role-to-location information and connection rules.

\noindent \textbf{Connection rules:} The connection rules are as follows and notice that except servers, the role list (role assignment) of switches can decide the only matrix for the correct connections among switches.

\begin{itemize}
\item Layer I: server, layer II: edge switches, layer III: aggregate switches, layer IV: core switches, only switches between neighbor layers have connections.
\item For core and aggregation role, if core's group id equals aggregation's index id, then connect, otherwise not.
\item For edge and aggregation role, if their group id are the same, then connect, otherwise not.
\item For server and edge role, if server's group id equals edge's index id, then connect, otherwise not.
\end{itemize}

\section{Topoconfig Algorithm}

Although we can not know the logical id of each devices as A3 does not need to record the device-to-location
information, we design an Topoconfig algorithm to
compute the role information of each device. Using such role information, we can directly know the expected
neighbor devices connected to each device. Then no matter malfunction detection or
fixation computation, address autoconfiguration can be done in P for FatTree with less malfunctions.

We first introduce the general idea
to solve the MGDP problem for FatTree. Then we design two exact algorithms
based on such idea, one can get the minimum fixation for FatTree with less than $k/2$ undirected link malfunctions and provide feasible fixation for FatTree with more malfunctions ($O(k^6)$).
The other is for FatTree with less than $k/4$ undirected link malfunctions ($O(k^3)$).

\begin{definition}
\textbf{x undirected link malfunctions} means the sum of both undirected link down and
added error is x.
\end{definition}

Using the role information of each node, we can directly know the expected neighbors connected to each node. The basic idea is that the switch role list can decides the only FatTree switch matrix
, using the role information in the last but one physical graph, we can get the last correct FatTree graph directly). Therefore, once we finding the switch-to-role mapping using the device graph matrix, we can get the fixation by comparing the FatTree switch matrix with the device graph matrix (the number of fixation steps equals to the  \textbf{Hamming distance} between those two matrixes). Then solving MGDP is transformed into finding the most suitable switch-to-role list to get the minimum fixation ($Optd(\pi)$).

\subsection{Exact algorithm 1}

We utilize the characteristics of FatTree to accelerate
role assignment.
Since the degree of server node is unlikely to be more than 1 physically, we do not consider the mistakes that a server is misused as a switch. The exact algorithm has 5 steps. In the first step, we strip out the $k^3/4$ least degree nodes which are servers. Second, with servers removed, the edges nodes have half degree as others. Thus we set the $k^2/2$ nodes with least degree as edge nodes. After determining the edge nodes, we classify the remaining nodes as aggregate nodes and core nodes by the number of edge nodes they are connected. Nodes have most connections with edge nodes are aggregate nodes. Then we group the edge nodes by the similarity of their direct neighbours and assign an id for each group. Next, we group core nodes the same as edge nodes and assign a random id to each group, too. Finally, we greedily assign a node to each aggregate role, by the number of common direct neighbours between the actual node and the role.

\begin{definition}
\textbf{Similarity} of two nodes $u$ and $v$ is defined as
  the number of the same neighbors connected to $u$ and $v$ directly, which can be computed by the inner product of $A(u)$ and $A(v)$. $A(u)$ is the row of $u$ in the adjacency matrix $A$ of graph $G$.
  \label{def:Similarity}
\end{definition}

\begin{figure}
\centering
\includegraphics [width=0.5\linewidth]{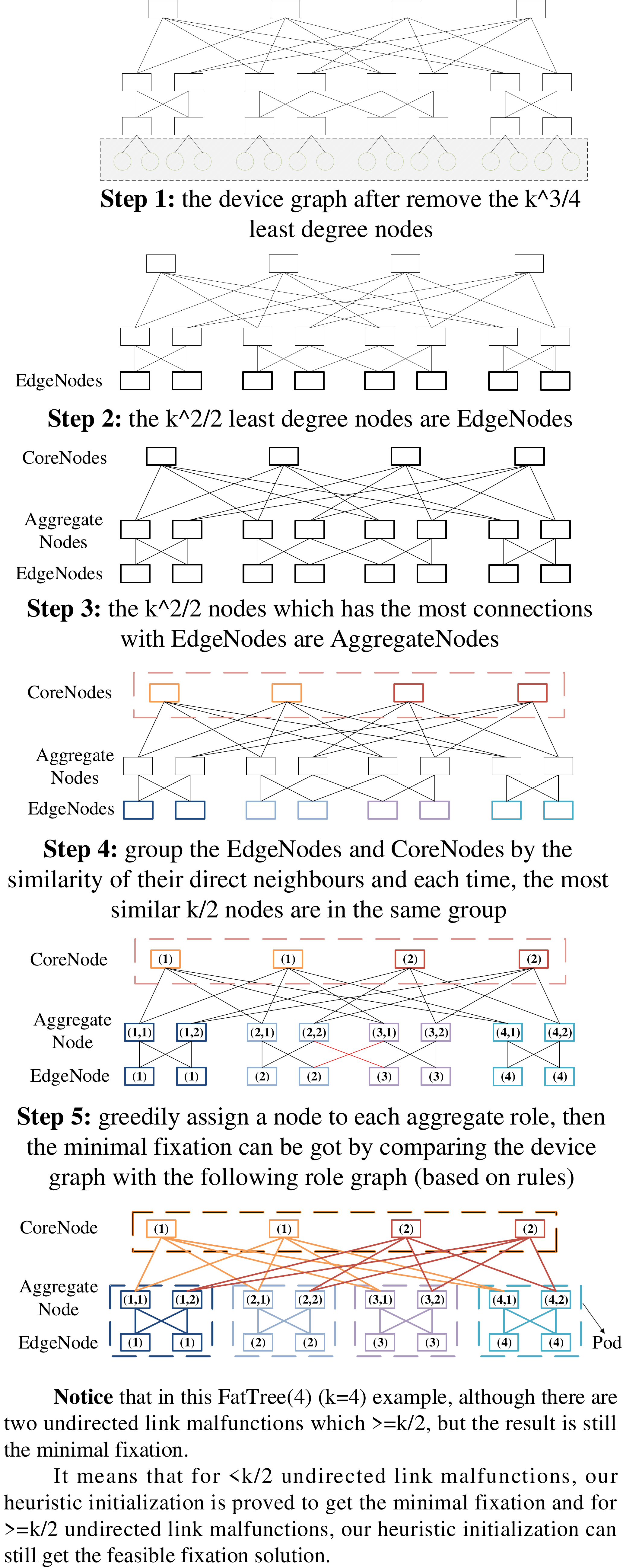}
\caption{Example of exact algorithm 1. }
\label{fig:example-heuristic-1}
\end{figure}

We will prove that for less than $k/2$ undirected link malfunctions, the output of our exact algorithm 1 is the minimum fix for any $k>4$
. (less than $k/2$ undirected link malfunctions means
the minimum fix step $\leq 2*(k/2-1)$, in other words if the output role list of exact algorithm 1 can generate $G_1$ and the deive graph is
$G_2$, then $d(\pi) = \sum_{(u,v)\in V_1\times V_1} |a_1(u,v) - a_2(\pi(u),\pi(v))| \leq k-2)$).

\begin{proof}

We will first prove that our exact algorithm initialization can assign the ``right'' role for each node step by step.
The basic idea is to prove it for the extreme condition (which means the correctness also holds for the condition of less malfunctions). We will  prove that within the bound, the role of each node will not be wrong using our algorithm.
The form is that for each step, we first introduce the conclusion, then we prove this conclusion holds even for the most extreme
malfunctions.

Step 1. the device graph after remove the $k^3/4$ least degree nodes includes all the switches. As our link malfunctions are less than $k/2$, the
server node can have at most $1+k/2-1=k/2$ edges (although this may not possible in the reality), but its degree is still less than other switch nodes which have
$k$ edges. The other extreme is that a switch node has more than $k-(k/2-1)=k/2+1$ edges, but its degree is still more than any server node.
A more general explanation is that the degree of any server can be $1+x$, and the degree of any switch can be $k-y$, $k-y-(1+x)=k-1-(x+y)>k/2-1>0$ (for
any $k \geq 4$)

Step 2. the $k^2/2$ least degree nodes are edge switches. Similarly, the degree of any non edge switch can be $k-x$ and the degree of any edge switch
can be $k/2+y$. Then due to link malfunctions are less than $k/2$, $x+y<k/2$. Therefore, $(k-x)-(k/2+y)=k/2-(x+y)>0$ which means the degree of any non
edge switch is larger than the degree of any edge switch in the device graph after remove the $k^3/4$ least degree nodes.

Step 3. the $k^2/2$ nodes which has the most connections with edge switches are aggregate switches and the rest $k^2/4$ nodes are core switches.
For any core switch with $x$ malfunction connections with edge switches and any aggregate switch with $k/2-y$ connections with edge switches,
$x-(k/2-y)=x+y-k/2<0$ which means the statement of step 3 holds.

Step 4. group the edge nodes and core nodes by the similarity of their direct neighbours and each time, the most similar $k/2$ nodes are in the same group. As we have proved that the level of each node is computed correctly, in this step, we consider a edge node with x links connected to the aggregate node in the other pod and $k/2-y$ links connected to its own pod, $k/2-y-x>k/2-(y+x)>0$ which means the similarity between this node and the other node in its own pod is larger than the similarity between it and the one in the other pod. For a core node, in the similar extreme condition, $k-y-x>k/2>0$, so step 4 is correct.

Step 5. greedily assign a node to each aggregate role. Similarly, if a aggregate node has x same connections with the other role and k-y connections same with it original role, as $k-y-x>k/2>0$, this node will be assigned to its original role which is correct.

Therefore, for any $k>4$ FatTree with less than $k/2$ undirected link malfunctions, the output of our exact algorithm 1 is the minimum fixation between the collected physical graph and the blueprint.

\end{proof}

\subsection{Exact algorithm 2}

Our idea is that after getting the physical graph collected from a DC,
we can first analyze the matrix of the physical graph using the
features of the DC topology to try to find the devices in each level.
For example, for FatTree, we try to figure out which switches are
edge switches, which are aggregation switches and which are core switches.
Then using the defination of roles (which is only related to the neighbors
of a node), we can group the nodes and mark the roles for those nodes
which have the ``correct'' neighbors. Next, we can also mark the roles for those nodes which have the ``incorrect'' links. Finally, the fixation
is the differences between the role-graph and the device graph. In summary, when we can not decides the roles of some nodes or the fixation result of exact algorithm 2 is larger than $2*(k/4-1)$, 
we know the number of undirected link malfunctions are larger than $k/4$, then the previous exact algorithm is triggerred or running parallely.
The following is the detail of this exact algorithm 2.

Inputting the device graph, and for less than $k/4$ undirected link malfunctions, minimal fixation can be obtained using the following steps.

1. Remove nodes whose degree $\leq k/4$ (should have $k^3/4$ nodes which are servers) to get the switch graph.

2. Hash each row of the switch graph.

\begin{figure}
	\centering
  \includegraphics [width=0.5\linewidth]{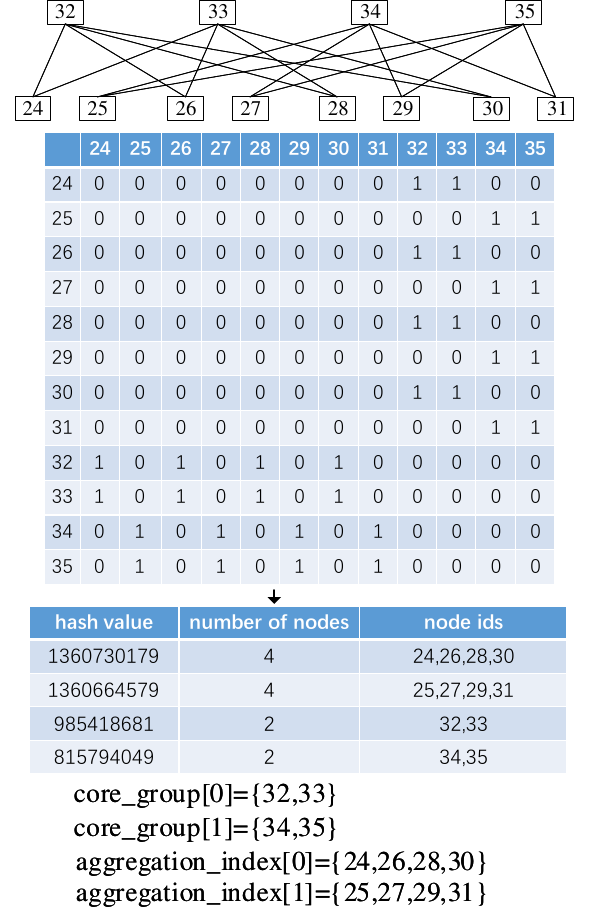}
	\caption{Example of the process of computing groups.}
	\label{fig:misauto}
	\end{figure}

3. Group the nodes whose corresponding rows have the same hash value. If a group has $k/2$ nodes, mark it as ``correct'' group, else mark it as
``incomplete'' group. An example is shown in Figure \ref{fig:misauto}.

4. Check the degree $x$ of any node in each group, if $x=k$, mark the group as ``core'' group. Else if $x=k/2$, mark the group
as ``edge'' group. Else, mark the group as ``malfunction'' group.

\begin{figure}
\centering
\includegraphics [width=0.5\linewidth]{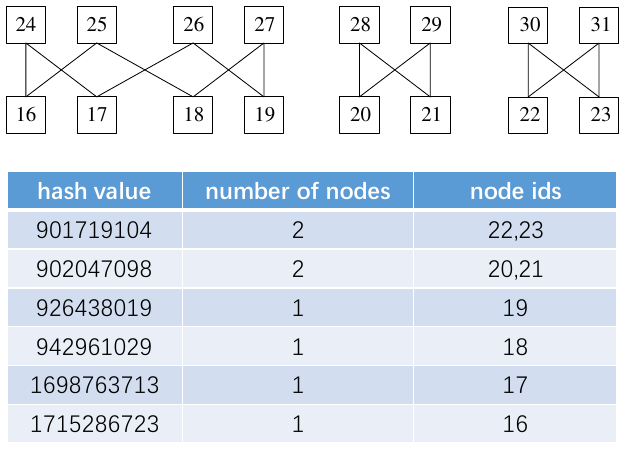}
\caption{An example with $\geq k/2$ undirected link malfunctions for $Fattree(k=4)$. The figure only shows the connections between aggregation switches and edge switches for simplicity.}
\label{fig:missautolocate}
\end{figure}

5. The union of nodes connected to the core groups and edge groups are aggregation switches. Specifically, we can mark the index of those nodes the same
as the core group id they connected to and the group of those nodes the same as the edge group id they connected to. Notice that for less than $k/4$ undirected link malfunctions, the number of edge group will be $k$ and the number of core group will be $k/2$ (including at most less than $k/4$ ``incomplete'' groups and the rest ones are ``incomplete malfunction'' groups).
Therefore, the union includes all the aggregation switches. Moreover, the nodes connected to the core groups includes all the aggregation switches and the nodes connected to the edge switches also includes all the aggregation switches, so the role, AggregateNode(group,index), of each aggregatation switch can be decided (including the malfunciton aggregate switches).

An example with $\geq k/2$ undirected link malfunctions is shown in Figure \ref{fig:missautolocate}. Notice that using our exact algorithm 2, we can find that the number of edge group is less than $k$ in this example, therefore exact algorithm 2 will output the information that the number of undirected link malfunctions is $\geq k/2$ and return in this step.

6. The nodes which have the unique hash value (except the upper marked aggregation switches) and the nodes belongs to the ``malfunction'' group
 are ``malfunctions"'. Using majority rule, we can decide their roles correctly for less than $k/4$ undirected link malfunctions. For each malfunction node,
 compute the \textbf{similarity} (inner product of corresponding rows, Definition \ref{def:Similarity})
 between it
 and
 any one node in the ``incomplete'' core or edge group and then find the most similar one
  (the group id of malfunction node is the same as the corresponding ``incomplete'' group id).

7. Minimal fixation can be obtained by computing the different 0,1 parts of the role-graph and the device graph.

8. For servers, we can check whether each edge switch connected to $k/2$ servers in the original device graph, if yes, all the servers are ``correct'',
else we can use the method proposed in section 4.4 of DAC \cite{ref:dac}, sequentially turn on the power of each rack to generate a record for
the mapping between MAC and rack. Then we can decide the suitable correction for servers.

Because there are all kinds of failure patterns, in DAC, they have four assumptions 1. most cases the malfunctions cause detectable node degree change; 2. miswiring
with no node degree change occurs after an administrator has checked the network and the degree-changing malfunctions have been fixed;
3. the number of nodes related to malfunctions is a considerably small amount over all the nodes; 4. $G_p$ and
$G_b$ have the same number of nodes and node degree patterns. We argue that the malfunctions can be fixed at one time to save the cost.

In our system, we try to analyze the physical data center topology under no assumptions, try to find out in what condition, we can locate the failures quickly.

\section{Evaluation}
\label{sec:evaluation}

\begin{table*}[h]
  \tiny
\caption{The scales of FatTree}\label{Tab:fatscale}
\centering
\begin{tabular}{|c|c|c|c|c|c|c|c|c|c|c|}
\hline
FatTree(n) & FatTree(20) & FatTree(30) & FatTree(40) & FatTree(50) & FatTree(60) & FatTree(70) & FatTree(80) & FatTree(90) & FatTree(100) \\
 \hline
Devices & 2500 & 7875 & 18000 & 34375 & 58500 & 91875 & 136000 & 192375 & 262500 \\
 \hline
Connections & 6000 & 20250 & 48000 & 93750 & 162000 & 257250 & 384000 & 546750 & 750000 \\
 \hline
\end{tabular}
\end{table*}

\begin{figure*}[!htbp]
\begin{subfigure}[b]{0.49\textwidth}
    \centering
\includegraphics[width=\linewidth]{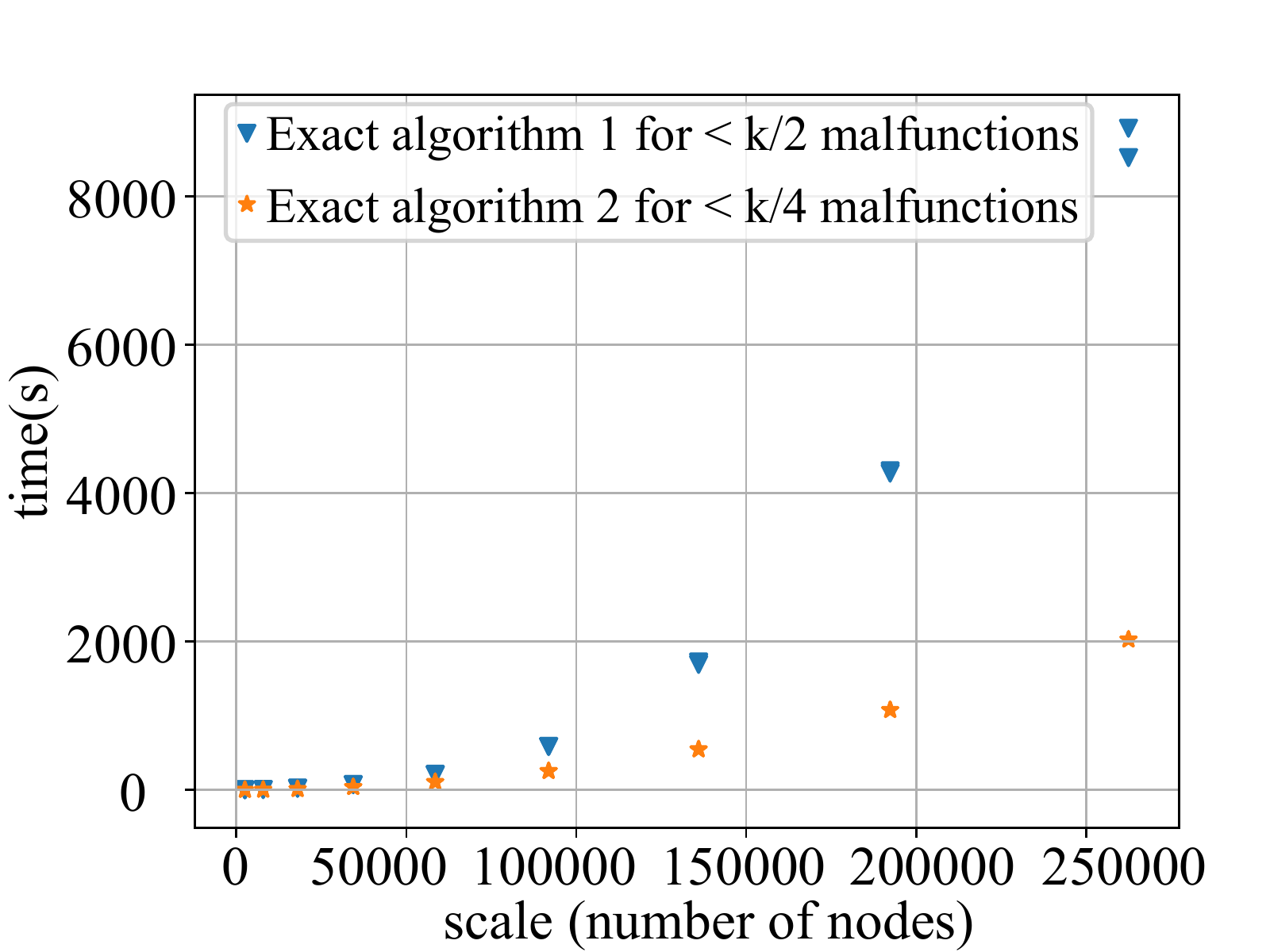}
\label{fig:compheuristicsk20-100legend}
\end{subfigure}
\hfill
\begin{subfigure}[b]{0.49\textwidth}
\centering
\includegraphics [width=\linewidth]{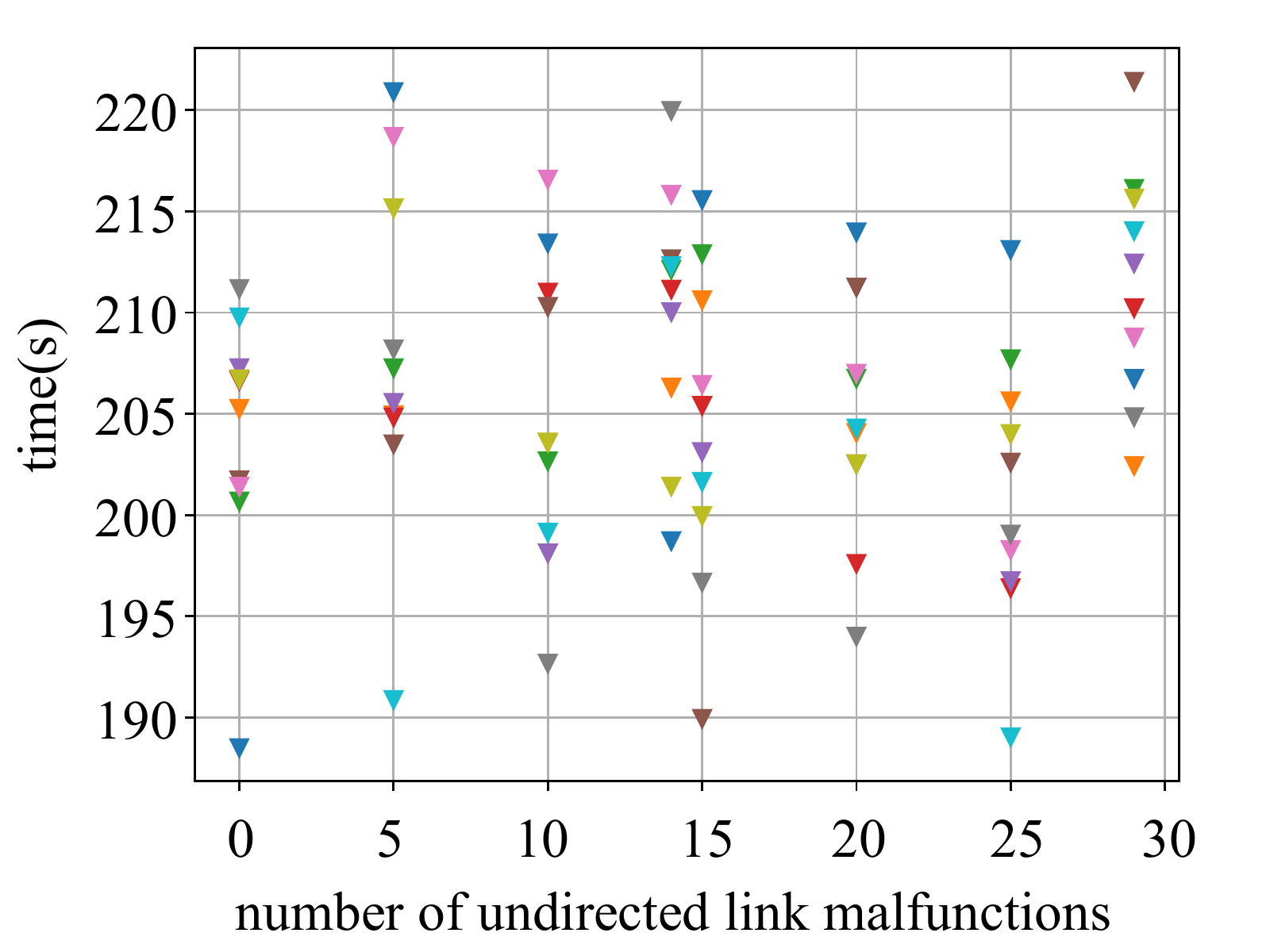}
\label{fig:exp5zsw}
\end{subfigure}
\vspace{-0.5cm}
\caption{(a) A3: Comparison between exact algorithm 1 and exact algorithm 2. (b) A3: Malfunction detection and fixation time vs. number of undirected link malfunctions for FatTree(k=60).
Figure shows computing time of exact algorithm 1 for each inputted physical graph with random malfunctions (10 physical graphs for each number of malfunctions: 0,5,10,14 (k/4-1),15,20,25,29 (k/2-1)).
}
\vspace{-0.5cm}
\end{figure*}

\vspace{-0.5cm}
\begin{figure*}[!htbp]
\begin{subfigure}[b]{0.49\textwidth}
\centering
\includegraphics [width=\linewidth]{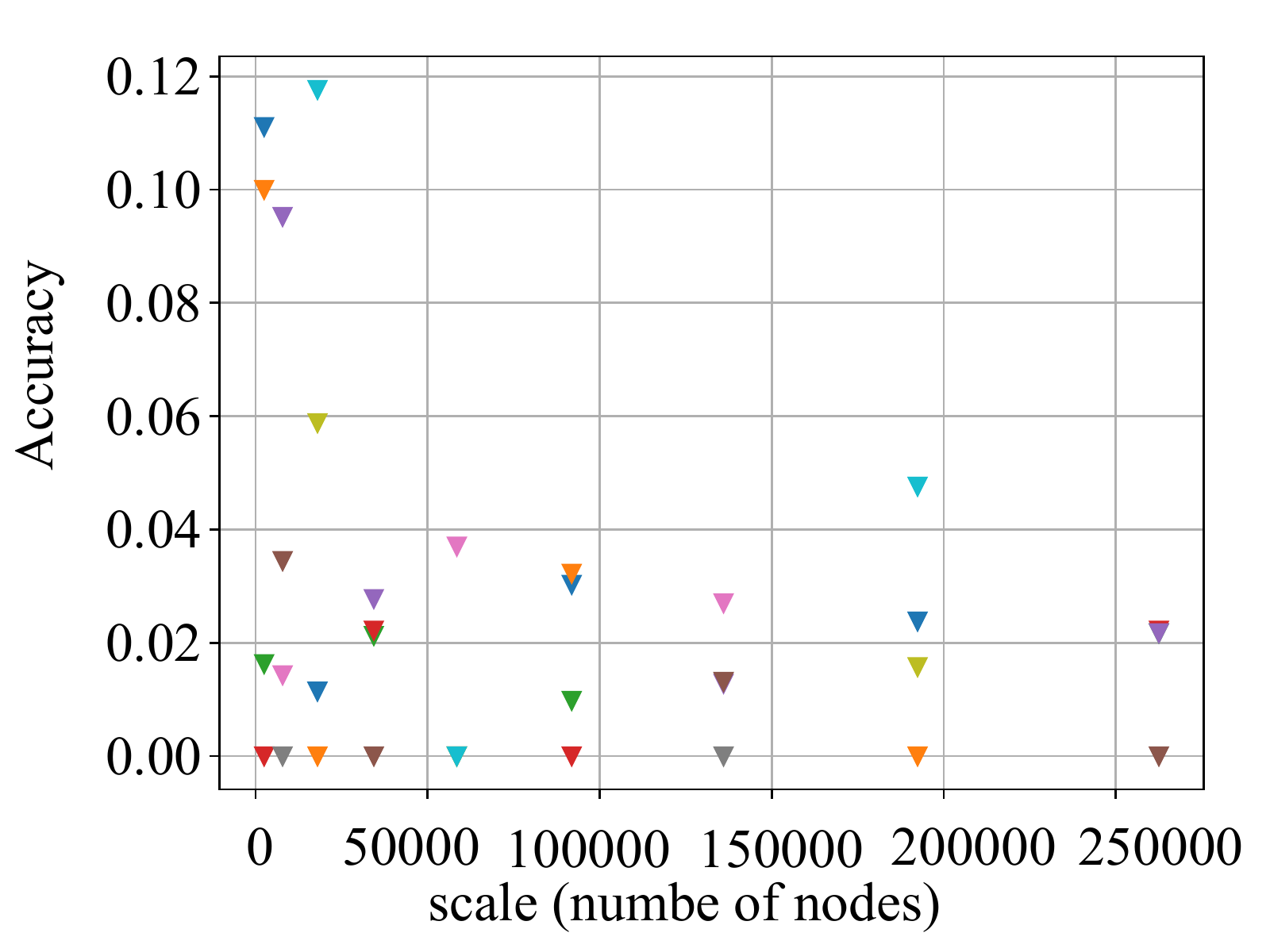}
\label{fig:dacAccdot015}
\end{subfigure}
\hfill
\begin{subfigure}[b]{0.49\textwidth}
\centering
\includegraphics [width=\linewidth]{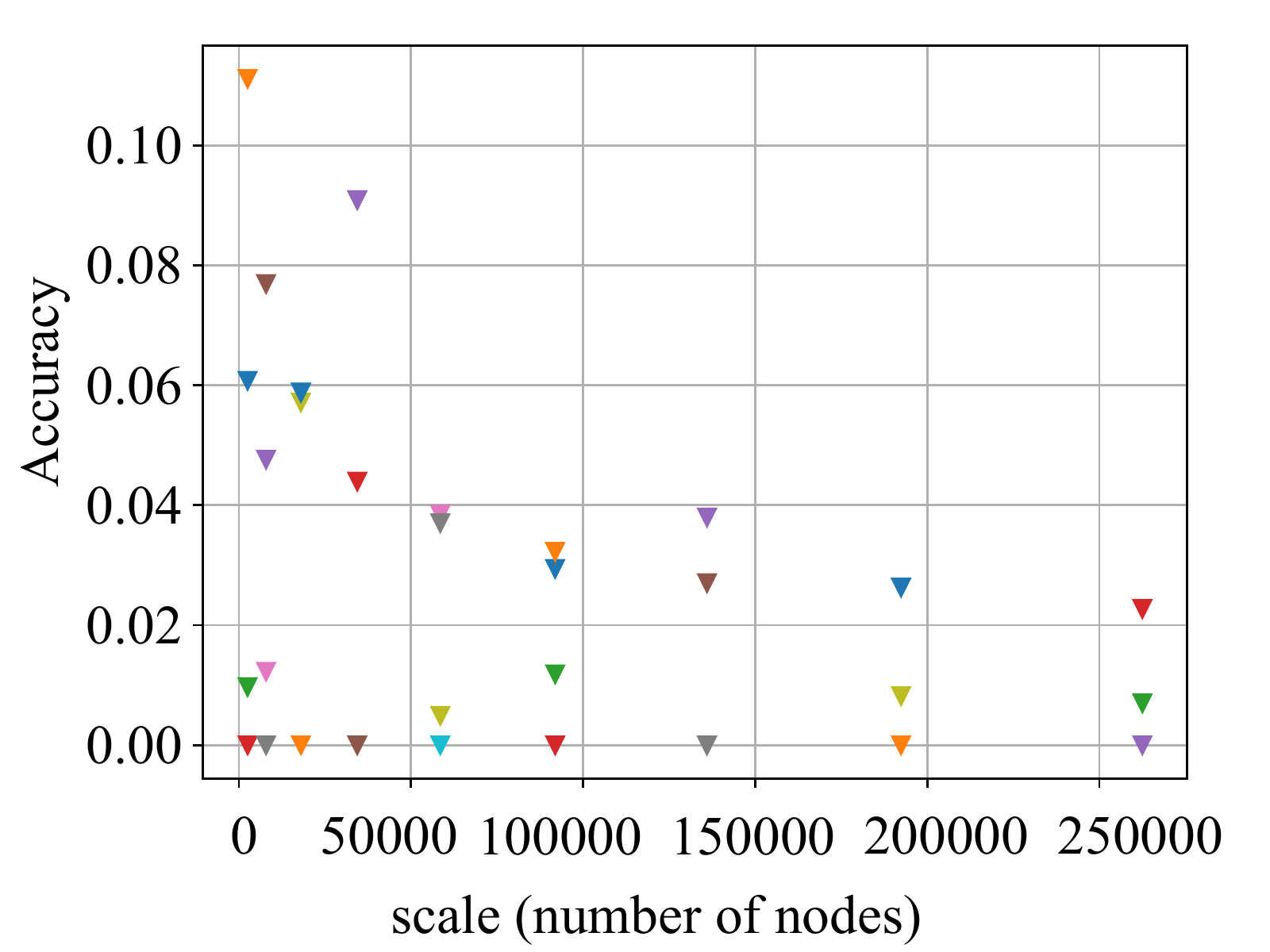}
\label{fig:dacAccdot03}
\end{subfigure}
\vspace{-0.5cm}
\caption{(a) and (b) are DAC's maximum accuracy for each number of malfunction of each scale (1.5\% and 3\% as anchor points).}
\vspace{-1cm}
\end{figure*}

\begin{figure*}[!htbp]
\begin{subfigure}[b]{0.49\textwidth}
  \centering
  \includegraphics [width=1.0\linewidth]{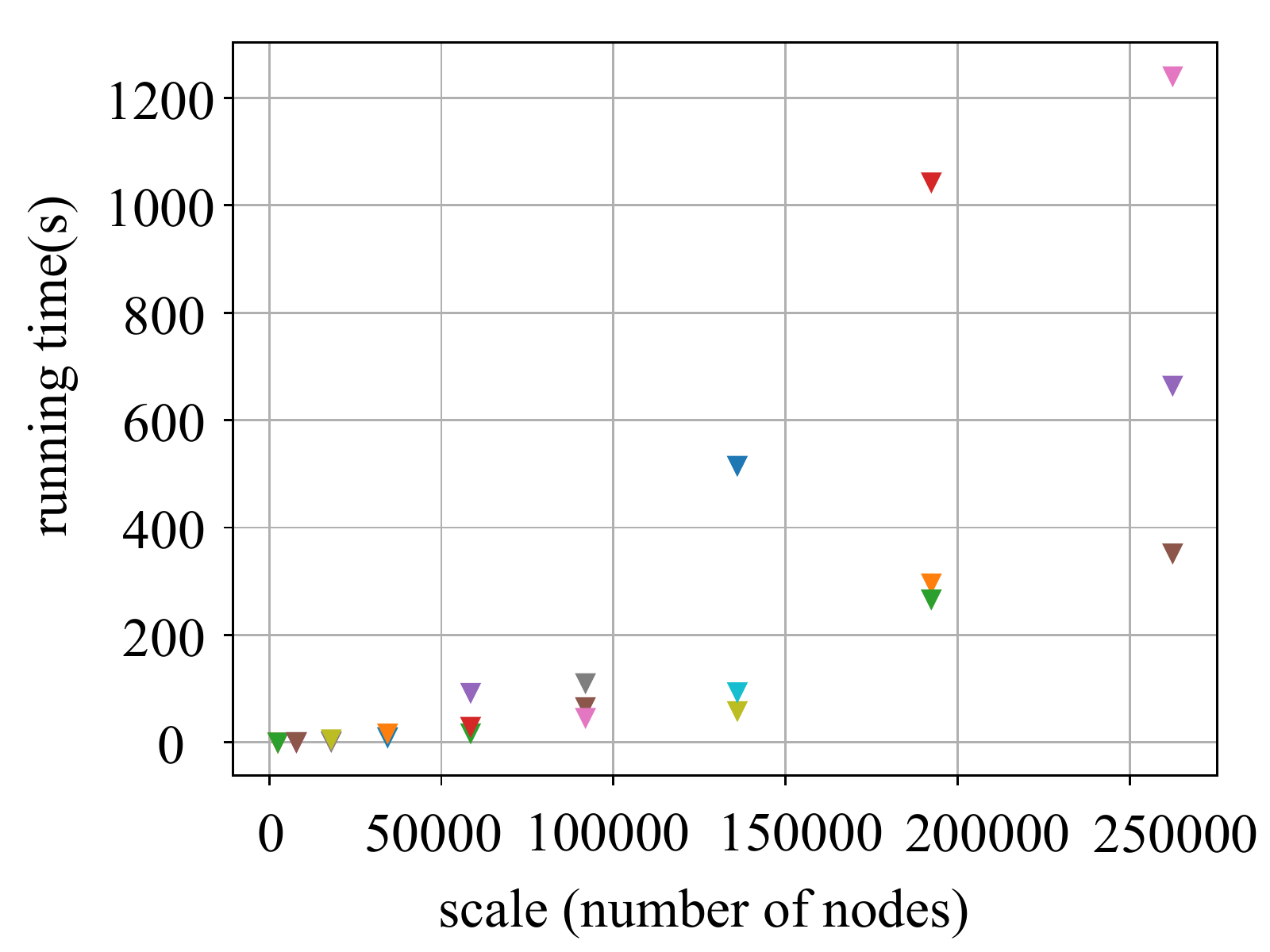}
  \label{fig:dacMalDetecttimedot015}
\end{subfigure}
\begin{subfigure}[b]{0.49\textwidth}
    \centering
  \includegraphics [width=1.0\linewidth]{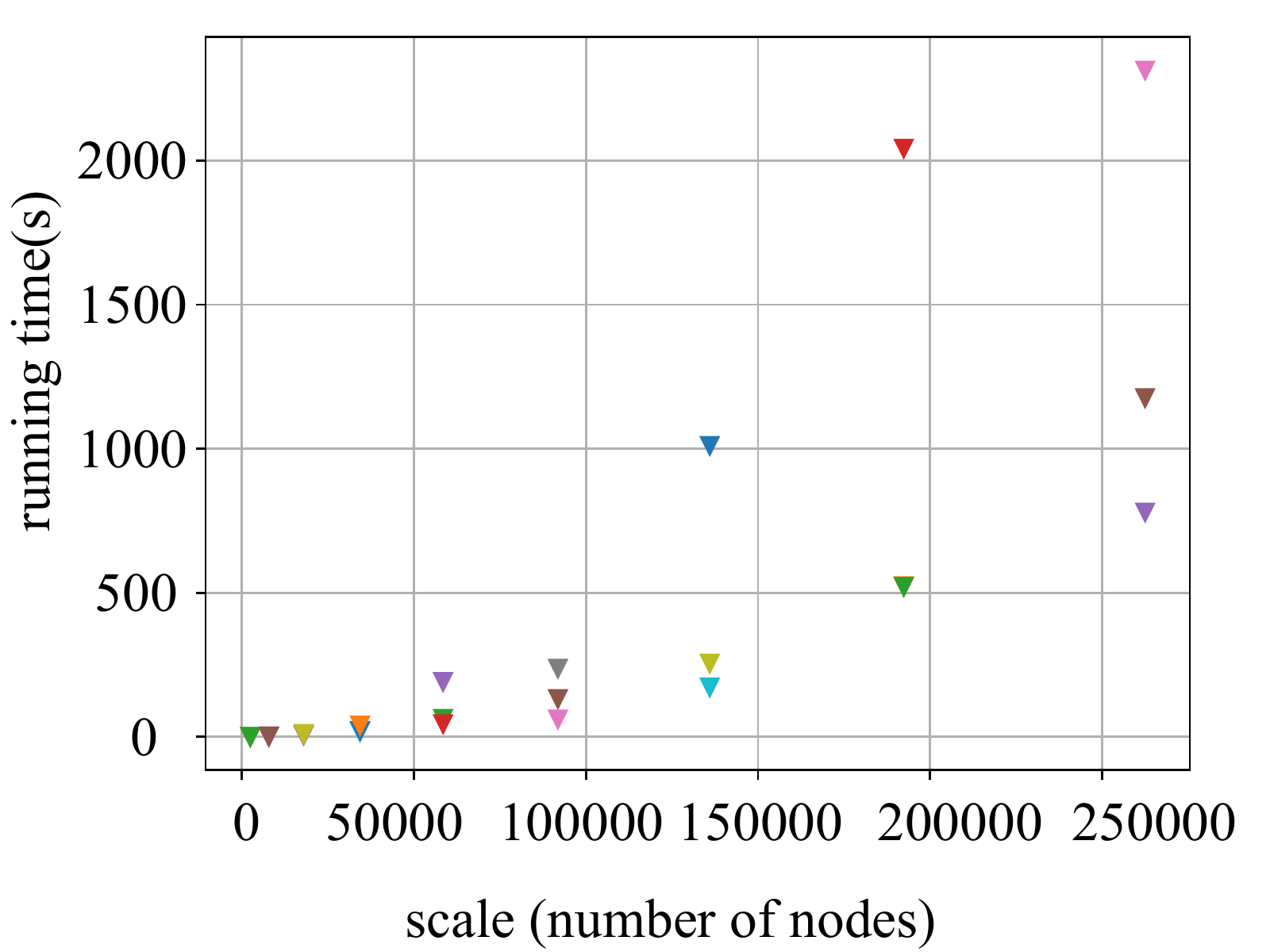}
  \label{fig:dacMalDetecttimedot03}
\end{subfigure}
\vspace{-0.5cm}
\caption{DAC: The average malfunction detection time for each number of malfunction of each scale (selecting 1.5\% and 3\% nodes as anchor points for (a) and (b) and remove the time of pre-computing SPLD of blueprint as suggested in DAC \cite{ref:dac}).}
\vspace{-1cm}
\end{figure*}

\subsection{Settings}

We evaluate A3 on FatTree
and give different test settings,
varying from network scale and the number of malfunctions.
Firstly, we measure the speed of A3 for different scale of FatTree,
including malfunction detection and malfunction correction (minimal fixation).
This metric is to show the scalability and how efficient A3 is.
Then, we change the number of malfunctions and calculate the CPU time overhead for FatTree(60), which shows the impact of malfunction quantities.
There are two metrics in our evaluation, which are the CPU calculation time and the accuracy.
For each number of malfunctions of each FatTree scale, we generate 5 random physical graph with that number of malfunctions. We run the experiments and collect the data on a server
which has a Xeon 3.40Hz 8-core CPU, 8GB DRAM, 1TB disk.

\subsection{Performance for A3's exact algorithms}

\noindent \textbf{Performance for scales.} As shown in Figure 7a
, actually for the same FatTree(k) scale, there are four average time of finding minimal fixation corresponding to four
types of number of malfunctions ($0,k/4-1,k/2-1,k/2$) for exact algorithm 1 and two ($0,k/4-1$) for exact algorithm 2. For each exact algorithm, we can see that average time almost not changes (coincidence with each other for different number of malfunctions) with the number of malfunctions for each scale. From the time increasing trend of each exact algorithm as the increasing of FatTree scale, we can find it meets the time complexity of each exact algorithm.

As we have introduced
the design section,
A3 solved MGDP (proved to be NP-complete for general graph)
in $O(k^6)$ and $O(k^3)$ for FatTree with any less than $k/2$ (exact algorithm 1) and $k/4$ (exact algorithm 2) undirected link malfunctions respectively.
Within each exact algorithm's malfunction bound, its accuracy is 100\% and beyond its bound, in our experiment, exact algorithm 1 still has accuracy of about 100\%.

\noindent \textbf{Performance for the number of malfunctions.} As shown in Figure 7b
, we can see that for FatTree(60), the time of finding minimal fixation for each random malfunction using
exact algorithm 1 for our Topoconfig algorithm is slightly changed for the same number of malfunctions. However, just like in Figure 7a,
we can notice that, the average time is almost not change as the increase of number of malfunctions.

\subsection{Performance for DAC's malfunction detection}

The number of malfunction in DAC's experiments is $k/4-1,k/2-1,k/2$.

\noindent \textbf{Accuracy.} As shown in Figure 8a and 8b
, the maximum accuracy of DAC's malfunction detection algorithm is still below 12\%. It means administrator may need large amount of time to check whether DAC's reported malfunctions are real malfunctions. 

\noindent \textbf{Malfunction detection time.} As shown in Figure 9a and 9b
, comparising with Figure 7a
, DAC's malfunction detection time is less than A3. It is because the number of anchor points is less and we use faster\_could\_be\_isomorphic(G1,G2) function in NetworkX of Python as the graph isomorphism algorithm to replace O2Mapping in the malfunction detection algorithm. The reason we use that function is 1. we try to implement DAC's O2Mapping but it is so difficult that we haven't debugged it well spending more than two weeks; 2. we try to use the is\_isomorphic(G1,G2) function in NetworkX to implement DAC's malfunction detection algorithm but sometimes, the computing time is too long (hours).

As to malfunctions without node degree changes, the idea of DAC is that they first find several nodes that ought to be corresponding between $G_p$ and $G_b$, then utilize those nodes as anchor points ($v/v'$s) to work over whether the subgraphs deduced from those two graphs are isomorphic ($x$-hop). The reasons of such low accuracy are 1. the malfunctions are randomly generated which include both miswirings with and without node degree changes; 2. DAC assumes some miswirings happened between $x$-hop and $(x+1)$-hop away from $v$ and the nodes in these two hops are suspicious. However, for miswirings include more than two links or malfunctions with node degree change beyond $(x+1)$-hop etc., this assumption does not hold; 3. DAC's suspicious node in $x$-hop and $(x+1)$-hop away from $v$ may include large amount of correct nodes, therefore the output list may not be accurate if the anchor points are not selected appropriately.

\subsection{Assumptions}
We focus on finding the minimum fixation for the collected physical network topology with less link malfunctions (e.g., link failures, cabling errors, miswirings do not cause any node degree changes \cite{ref:dac}, etc.). For the limitation that we assume that any two switches can be connected with the same cost,
we argue that it is probably not to break when we find the minimum fixation. Connecting switches in different racks happens if and only if there are already large amount of connections between them in the physical graph which can be noticed when network builders connect them or they are just right connections with such long distance.

\section{Related Work}
\label{sec:related}

The authors of DAC~\cite{ref:dac} propose a generic and automatic address configuration structure for DCN. It first generates blueprint which defines the connections of switches and servers labeled by logical IDs automatically and learns the physical graph with device IDs. Then it runs the logical-to-device ID mapping by using the theory of graph isomorphism.
Malfunctions are not uncommon because the large scale of a DC. When there are malfunctions, DAC needs to wait until all the malfunctions are corrected.

Therefore, ETAC~\cite{ref:etac}, is proposed to configure DC addresses in the presence of malfunctions. It first converts the physical topology into a conceptual graph and remove all the malfunction machines and all their links from the graph to mathematically transform the problem of address configuration into the problem of induced subgraph isomorphism.

DAC and ETAC both use the method proposed in DAC to find malfunctions. As mentioned in DAC, malfunction detection problem can be presented as follows. Given $G_b$(blueprint) and $G_p$(physical graph collected from a data center), the problem to locate all the malfunctioning parts in the graph $G_p$ is equal to find  the maximum common subgraph (MCS~\cite{ref:mcs}) $G_{mcs}$ of $G_b$ and $G_p$. Thus, the malfunctioning parts are the differences between $G_{mcs}$ and $G_p$. All machines (i.e.,servers or switches) involved in these parts, called malfunctioning machines, can be detected. But the MCS problem has been proven to be NP-complete 
and APX-hard~\cite{ref:mcsapx}. As to malfunctions without node degree change, the idea of DAC is that they first find several nodes that ought to be corresponding between $G_p$ and $G_b$, then utilize those nodes as anchor points to work over whether the subgraphs deduced from those two graphs are isomorphic. However, in the case that the anchor points are not selected correctly, it will be very time-consuming, although it use shortest path length distribution (SPLD)~\cite{ref:dac} to select the anchor points.

Different from graph isomorphism algorithm designed by DAC and induced subgraph isomorphism algorithm proposed by ETAC which are
designed for general graph, we utilize the feature of each data center topology like FatTree to automatically configure its devices' addresses quickly
even in the presence of malfunctions. Moreover, we can compute the minimal fixation to direct builders to repair malfunctions efficiently.

Weaknesses of previous works are summarized below.

\noindent \textbf{Data Center Slowly Building Process}.
As previous blueprint~\cite{ref:dac} requires information
about interconnections between devices, builders need to record the device-to-location information in order use such blueprint which is a huge work and
such record may be error-prone.

\noindent \textbf{Not Support Incremental Deployment and real time malfunction detection}. Previous blueprint does not include the  topological properties which are easy to follow and check each connection. The network managers then can not detect the mis-cable malfuncitons at real time although they can check whether the connection works well.

\noindent \textbf{Inaccurate Fault Location and No Fix Suggestion}. The accuracy of previous scheme depends on the number of anchor points they selected for detection versus the number of miswirings in the network.
Although previous work claims they can always conservatively select a larger percentage of anchor points to start their detection and most likely they
will detect all miswirings, it will make the process very time-consuming and there is no proof for the guarantee of accuracy.

Due to inaccurate fault location and no fix suggestion, maintenance staff needs to first check which devices are really malfunctions. Then they need to figure out how to fix those malfunctions and fix them manually. This process is very time-consuming and error-prone.

\section{Conclusion}
\label{sec:conclusion}
In this paper, we have designed, evaluated and implemented A3, a topology aware automatic address configuration and malfunction fixation system.
To the best of our knowledge, this is the first work to introduce roles and their connection rules to the blueprint and using such topology aware
information to accelerate automatic address configuration and malfunction detection.
Moreover, this is also the first work to compute the minimum fixation for data center with malfunctions. At the core of A3 is its device-to-role
mapping. A3 has made an innovation in abstracting the problem of finding minimal fixation to the minimum graph difference problem (proved to be NP-complete)
and
solved it in $O(k^6)$ and $O(k^3)$ for any less than $k/2$ and $k/4$ undirected link malfunctions respectively using the roles and rules for FatTree.

{\small
\bibliographystyle{abbrv}
\balance
\bibliography{main}

\begin{thebibliography}{10}

\bibitem{ref:fattree}
M.~Al-Fares, A.~Loukissas, and A.~Vahdat.
\newblock A scalable, commodity data center network architecture.
\newblock In {\em Proceedings of the ACM SIGCOMM 2008 conference on Data
  communication}, SIGCOMM '08, pages 63--74, New York, NY, USA, 2008. ACM.

\bibitem{ref:jupiter}
{Arjun Singh and Joon Ong and Amit Agarwal and Glen Anderson and Ashby
  Armistead and Roy Bannon and Seb Boving and Gaurav Desai and Bob Felderman
  and Paulie Germano and Anand Kanagala and Jeff Provost and Jason Simmons and
  Eiichi Tanda and Jim Wanderer and Urs H\"olzle and Stephen Stuart and Amin
  Vahdat}.
\newblock Jupiter rising: A decade of clos topologies and centralized control
  in google's datacenter network.
\newblock In {\em Sigcomm '15}, 2015.

\bibitem{ref:mces2012}
L.~Bahiense, G.~Manić, B.~Piva, and C.~Souza.
\newblock The maximum common edge subgraph problem: A polyhedral investigation.
\newblock {\em Discrete Applied Mathematics}, 160:2523–2541, 12 2012.

\bibitem{ref:mcs}
L.~Bahiense, B.~Piva, and C.~C.~D. Souza.
\newblock The maximum common edge subgraph problem: A polyhedral investigation.
\newblock {\em Discrete Applied Mathematics}, 160(18):2523--2541, 2012.

\bibitem{ref:ceci}
B.~Bhattarai, H.~Liu, and H.~H. Huang.
\newblock {CECI:} compact embedding cluster index for scalable subgraph
  matching.
\newblock In {\em Proceedings of the 2019 International Conference on
  Management of Data, {SIGMOD} Conference 2019, Amsterdam, The Netherlands,
  June 30 - July 5, 2019.}, pages 1447--1462, 2019.

\bibitem{ref:dac}
K.~Chen, C.~Guo, H.~Wu, J.~Yuan, Z.~Feng, Y.~Chen, S.~Lu, and W.~Wu.
\newblock Generic and automatic address configuration for data center networks.
\newblock In {\em {SIGCOMM}}, 2010.

\bibitem{ref:ruben}
R.~J.~S. García.
\newblock Exploiting symmetry in network analysis, 2018.

\bibitem{ref:vl2}
A.~Greenberg, N.~Jain, S.~Kandula, C.~Kim, P.~Lahiri, D.~Maltz, P.~Patel, and
  S.~Sengupta.
\newblock Vl2: A scalable and flexible data center network.
\newblock In {\em {SIGCOMM}}, 2009.

\bibitem{ref:subisodp}
M.~Han, H.~Kim, G.~Gu, K.~Park, and W.~Han.
\newblock Efficient subgraph matching: Harmonizing dynamic programming,
  adaptive matching order, and failing set together.
\newblock In {\em Proceedings of the 2019 International Conference on
  Management of Data, {SIGMOD} Conference 2019, Amsterdam, The Netherlands,
  June 30 - July 5, 2019.}, pages 1429--1446, 2019.

\bibitem{ref:between}
R.~Hoffmann, C.~McCreesh, and C.~Reilly.
\newblock Between subgraph isomorphism and maximum common subgraph.
\newblock In {\em Proceedings of the Thirty-First {AAAI} Conference on
  Artificial Intelligence, February 4-9, 2017, San Francisco, California,
  {USA.}}, pages 3907--3914, 2017.

\bibitem{ref:asap}
A.~P. Iyer, Z.~Liu, X.~Jin, S.~Venkataraman, V.~Braverman, and I.~Stoica.
\newblock {ASAP:} fast, approximate graph pattern mining at scale.
\newblock In {\em 13th {USENIX} Symposium on Operating Systems Design and
  Implementation, {OSDI} 2018, Carlsbad, CA, USA, October 8-10, 2018.}, pages
  745--761, 2018.

\bibitem{ref:mcsapx}
V.~Kann.
\newblock On the approximability of the maximum common subgraph problem.
\newblock In {\em Proceedings of the 9th Annual Symposium on Theoretical
  Aspects of Computer Science}, STACS '92, pages 377--388, London, UK, UK,
  1992. Springer-Verlag.

\bibitem{ref:etac}
X.~Ma, C.~Hu, K.~Chen, C.~Zhang, H.~Zhang, K.~Zheng, Y.~Chen, and X.~Sun.
\newblock Error tolerant address configuration for data center networks with
  malfunctioning devices.
\newblock In {\em Proceedings of the 2012 IEEE 32nd International Conference on
  Distributed Computing Systems}, ICDCS '12, pages 708--717, Washington, DC,
  USA, 2012. IEEE Computer Society.

\bibitem{ref:mcs2017}
C.~McCreesh, P.~Prosser, and J.~Trimble.
\newblock A partitioning algorithm for maximum common subgraph problems.
\newblock pages 712--719, 08 2017.

\bibitem{ref:cheiccc}
C.~Zhang, H.~Xu, and C.~Hu.
\newblock Performance impact inference with failures in data center networks.
\newblock In {\em 2016 IEEE/CIC International Conference on Communications in
  China (ICCC)}, pages 1--6, July 2016.

\bibitem{ref:a3poster}
C.~Zhang, S.~Zhang, B.~Jin, W.~Li, Z.~Wang, and Y.~Wang.
\newblock A3: An automatic malfunction detection and fixation system in fattree
  data center networks.
\newblock In {\em Proceedings of the ACM SIGCOMM 2019 Conference Posters and
  Demos}, SIGCOMM Posters and Demos '19, pages 24--26, New York, NY, USA, 2019.
  ACM.

\end{thebibliography}
}

\end{document}